\documentclass[conference]{IEEEtran}

\usepackage{graphics} 
\usepackage{epsfig} 
\usepackage{mathptmx} 
\usepackage{times} 
\usepackage{amsmath} 
\usepackage{amssymb}  
\usepackage{color}
\usepackage{listings}
\usepackage{subfigure}
\usepackage{hyperref}
\usepackage{verbatim}

\title{\LARGE \bf
Deep-Learnt Classification of Light Curves
}

\begin{document}
\bstctlcite{IEEEexample:BSTcontrol}
\author{\IEEEauthorblockN{A Mahabal\IEEEauthorrefmark{1},
K Sheth\IEEEauthorrefmark{2},
F Gieseke\IEEEauthorrefmark{3}, 
A Pai\IEEEauthorrefmark{3},
S G Djorgovski\IEEEauthorrefmark{1},
A J Drake\IEEEauthorrefmark{4},
M J Graham\IEEEauthorrefmark{1}, and
CSS/CRTS/PTF Teams
}
\IEEEauthorblockA{\IEEEauthorrefmark{1}Center for Data-Driven Discovery, California Institute of Technology, Pasadena, CA, 91125}
\IEEEauthorblockA{\IEEEauthorrefmark{2}Indian Institute of Technology Gandhinagar, Palaj, Gandhinagar, 382355, India}
\IEEEauthorblockA{\IEEEauthorrefmark{3}Department of Computer Science, University of Copenhagen, Copenhagen, Denmark}
\IEEEauthorblockA{\IEEEauthorrefmark{4}Cahill Center for Astronomy and Astrophysics, California Institute of Technology, Pasadena, CA, 91125}
E-mails: 
\texttt{aam@astro.caltech.edu},
\texttt{kshiteej.sheth@iitgn.ac.in},
\texttt{fabian.gieseke@di.ku.dk},\\
\texttt{akshay@di.ku.dk},
\texttt{george@astro.caltech.edu},
\texttt{ajd@astro.caltech.edu},
\texttt{mjg@caltech.edu}
}

\maketitle
\begin{abstract}

Astronomy light curves are sparse, gappy, and heteroscedastic. As a result standard time series methods regularly used for financial and similar datasets are of little help and astronomers are usually left to their own instruments and techniques to classify light curves. A common approach is to derive statistical features from the time series and to use machine learning methods, generally supervised, to separate objects into a few of the standard classes. 
In this work, we transform the time series to two-dimensional light curve representations in order to classify them using modern deep learning techniques. In particular, we show that convolutional neural networks based classifiers work well for broad characterization and classification.
We use labeled datasets of periodic variables from CRTS survey and show how this opens doors for a quick classification of diverse classes with several possible exciting extensions.

\end{abstract}

\section{Introduction}
\label{sec:intro}

Astronomy has always boasted of big datasets. The data holdings are getting even larger due to surveys that observe hundreds of millions of sources hundreds of time. The observations are a time series of flux measurements called light curves. The staple for discovery has been the flux variations of individual astronomical objects as noted through such light curves - that is where the science is. The large irregular gaps in observing cadence makes  classification challenging.  Traditionally statistical features have been derived from the light curves in order to do follow-up classification (see, e.g., \cite{Richards2011,Donalek2013,Graham2014}). The features include standard statistical measures like median, skew, kurtosis as well as specialized domain knowledge based ones such as `\textit{fading profile of a single peaked fast transient}'. The standard features do not carry special powers for classifying a varied set of objects. The designer features are better for specific classes, but carry with them a bias that does not necessarily translate to the classification of a wider set.

In \cite{Mahabal2011} we introduced a two-dimensional mapping of the light curves based on the changes in magnitude ($dm$) over the available time-differences ($dt$). In this work, we mold the $dm-dt$ mapping into an image format that is suitable as input for \emph{convolutional neural networks} (CNNs or ConvNets)~\cite{LeCun2015}. By bringing to bear the machinery of CNNs we are able to conjure a large number of features unimagined so far. We use labeled sets to train the CNN as a classifier and following validation we classify  light curves from the Catalina Real-Time Transient Survey (CRTS; \cite{Djorgovski2011,Drake2009,Mahabal2011,Mahabal2012,Djorgovski2016}).

\section{Data}
\label{sec:data}
The three CRTS surveys span 33,000 sq. degrees encompassing light curves of close to half a billion sources. Of these, the 0.7m CSS telescope yields $\sim150$ million light curves. The light curves span well over ten years, and are homogeneous in that all are collected using white-light without a filter, and with an asteroid-searching cadence of four images in 30 minutes. As is typical of the astronomical objects in wide-area surveys a vast majority of these sources ($>90\%$) are non-variable during the survey life-time and within the typical $\sim0.1$ mag error-bars for CSS. The remaining sources - variables - can be broadly classified as periodic and stochastic. The irregularly spaced sparse light curves mean that often even the periodically variable sources do not seem obviously so. There is a third category, that of transients like supernovae and flaring stars which exhibit enhanced activity over a short period and otherwise a quiescent and relatively flat (within error-bars) light curve.

Getting a large, uniform, well-labelled dataset is a challenge in itself. In order to keep the problem simple during early experiments, we start with a sample of $\sim50k$ periodic variables from the CRTS North (CRTS-N) survey [8]. There are 17 classes represented in the sample. Ten of these have fewer than 500 members. We exclude them from our experiments for now and will include them in future studies. The numbers for the remaining seven classes are given in Table~\ref{tab:members}. These classifications have been carried out by humans mostly based on calibrated light curves, their phased versions after periods were determined, and some auxiliary information on the objects. A little over $10\%$ of these have spectroscopic confirmation of the exact classification. As a result some misclassifications, especially in a nearby class can not be ruled out, especially for objects that are fainter and/or have fewer observations.

\begin{figure}
\centering
	\includegraphics[width=\columnwidth]{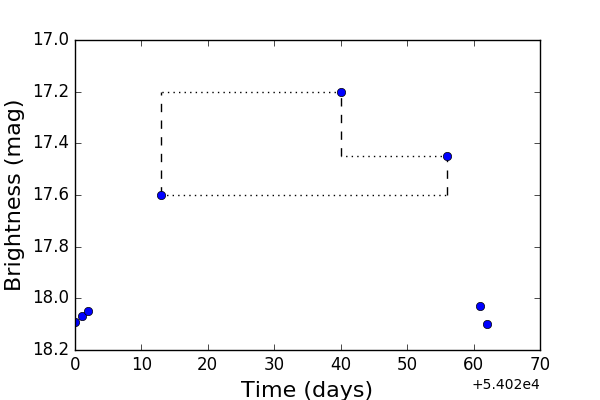}
    \caption{Part of a light curve is shown without error-bars to demonstrate $dm$ (dashed lines) and $dt$ (dotted lines) values. Each pair of points in the light curve leads to one $dmdt$ pair. Three pairs are shown. These then populate the $dmdt$ grid of Fig.~\ref{fig:dmdtgrid} and make the 23x24 $dmdt$-images in Figs.~\ref{fig:dmdt},\ref{fig:classback} etc.}
    \label{fig:dmdtschem}
\end{figure}

\section{DMDT Mappings}
\label{sec:technique}

A light curve consists of brightness variations as a function of time. Besides the time (expressed here in days - MJD), and brightness (expressed here in the traditional inverse logarithmic scale - mags), we also have information about the error in magnitudes. 

\begin{figure}
	\includegraphics[width=\columnwidth]{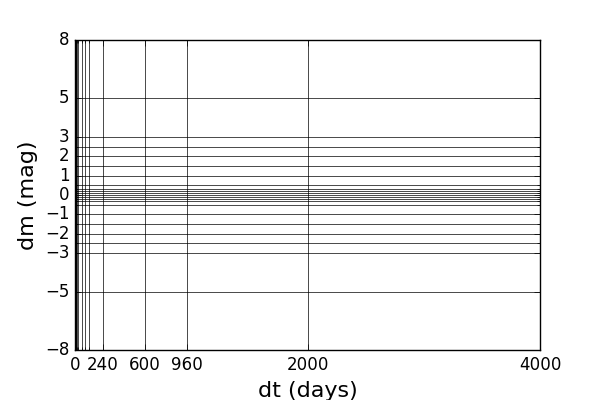}
    \caption{The $dmdt$ grid associated with our fiducial $dm$ and $dt$ spacings. Most labels near dt=0 and dm=0 are not printed due to crowding (see Table~\ref{tab:dmdtbins} for the full list). Each unequal-area rectangle here translates to one of the equal-area pixels in the 23x24 images used with CNNs (e.g. Figs.~\ref{fig:dmdt} and \ref{fig:classback}).}
    \label{fig:dmdtgrid}
\end{figure}

For each pair of points in a light curve we determine the difference in magnitude ($dm$) and the difference in time ($dt$). This gives us $p = \binom{n}{2} = n*(n-1)/2$ points for a light curve of length $n$ (see Fig.~\ref{fig:dmdtschem}). These points are then binned for a range of $dm$ and $dt$ values. The resulting binned $2D$ representation is our $2D$ mapping from the light curve. 
The bin boundaries we have used are:
$dm = \pm[0,0.1,0.2,0.3,0.5,1,1.5,2,2.5,3,5,8]$ mags and 
$dt = [1/145,2/145,3/145,4/145,1/25,2/25,3/25,1.5,2.5,3.5,\\4.5,5.5,7,10,20,30,60,90,120,240,600,960,2000,4000]$ days.
The $23\times24$ bins are in approximately a semi-logarithmic fashion so that frequent small magnitude changes are distributed over many bins, and the infrequent large magnitude changes can be combined together (see Fig.~\ref{fig:dmdtgrid}). Similarly the more important rapid changes are well represented, and the slower changes are clubbed together. This is akin to histogram equalization, but not forced to be exact.  In case of $dt$ these take into consideration the CSS cadence of four images in 30 minutes. The image intensity, $i$, is normalized by $p$ to account for varying lengths of original light curves, and stretched to fit between 0 and 255: $i = (255*n_{\rm bin}/p + 0.99999)_{\rm int}$. Thus, a bin that does not include a single point, now has an intensity of zero, and a bin that had at least one point has an intensity of at least 1, and the bin that had the maximum number of points has an intensity of at most 255. 255 is reached only when all points are in one bin - the typical max values we have seen in different classes are around 50. Unnormalized values go to several thousand based on the length of the light curve. The normalization also ensures that we can use our data to fine-tune other pre-learned networks. For inclusion in the training set, we require that the light curves contain at least 20 points.

The $2D$ representations - called $dmdt$-images hereafter - reflect the underlying structure from variability of the source. The $dmdt$-images are translation independent as they consider only the differences in time. A light curve reflected about the x-axis will provide a $dmdt$-image reflected about the y-axis. Thus the structure above and below the $dm=0$ line is discriminating for sources. In a sense the $dmdt$-images are like a structure function without consideration to the error-bars. The error-bars tend to be heteroskedastic, but are broadly a function of magnitude, and unless an individual source varies a lot, tend to be similar. In particular, the way the $dmdt$-technique works, the error-bars for neighboring points in the mapped version are similar. Taking into consideration the error-bars would be equivalent to smoothing along the $y$-axis ($dm$). For now we ignore the error-bars (but see Section~\ref{sec:tests}). 

Since the $dmdt$-image is a straightforward mapping of a light curve and the CNN can bring out features hidden therein, the proposed method opens up at least two important avenues. (1) Implication for real-time classification of variables and transients: A sparse light curve of recently discovered object will correspond to a $dmdt$-image that is just a sparse version of the $dmdt$-image that would be formed from the corresponding non-sparse light curve. Since some of the unique features are accentuated at discovery, the discriminative power would already be encapsulated in the $dmdt$-image. (2) Transfer learning: Training based on $dmdt$-images from one survey can be used to classify $dmdt$-images from another survey. We demonstrate this on a set of variables from CRTS-S a Southern set corresponding to the main training set used here \cite{Drake2017}, and corresponding PTF data \cite{Law2009}.

There are three ways in which we experiment with the setup to improve performance: (1) Change the $dmdt$ bins for optimality - these can be done based on the survey cadence, or based on the classes being considered, (2) Change the layers of the CNN depending on number of classes, size of training sample, possible ways in which unbalancedness between the classes is remedied (or not), and (3) modifying the light curve to $dmdt$-image mapping to bring out features in the classes being separated. We look at these in the next two sections.


\begin{table}
	\centering
	\caption{Number of objects belonging to the seven classes that have at least 500 members. The variable types include EW (contact binaries), EA (detached binaries), three types of RR Lyrae; and Mira and semi-regulars lumped into LPV. RS CVn's are rotating variables. Thus broadly speaking we have three classes: binaries, pulsating, and rotating. {\it Class} refers to the numeric labels used in \cite{Drake2014}.}
	\label{tab:members}
	\begin{tabular}{lccccccr} 
\hline
Type & EW&EA &RRab &RRc &RRd & RSCVn&LPV\\
\hline
Class & 1&2 & 4 & 5 & 6&8&13 \\
Num 
&30743&4683&2420&5469&502&1522&512\\
\hline
	\end{tabular}
\end{table}

\section{Convolutional Networks}
\label{sec:arch}

Recently, so-called deep learning techniques have become very popular in machine learning and various application domains. This is, in particular, the case for \emph{convolutional neural networks}~(CNNs), which are a special type of \emph{artificial neural networks}~(ANN)~\cite{Hastie2008,Murphy2012}. In astronomy, CNNs have been used for a few problems where structure in images is obvious (e.g. \cite{Dieleman2015} uses them for classifying galaxies based on morphology and \cite{CabreraVives2016} for detecting supernovae). We provide brief description of ANNs and CNNs here. More details can be found in \cite{Hastie2008,Murphy2012,LeCun2015} etc. 

An ANN is based on several \emph{layers}.
The overall input data (e.g., images) are provided to the \emph{input layer}.
The output of one layer serves as input for the next.
The last layer forms the output of the network. For instance, in a multiclass setting, the output layer would contain one node per class. The layers between the input and the output layer are the \emph{hidden layers}. For a standard ANN, the layers are fully connected, with each node of one layer connected to all nodes of the next.

A CNN is an extension of classical ANNs and consists of different types of layers. As before, we have an input and an output layer. Further, the last layers before the output layer are often standard fully connected layers, called \emph{dense} layers. The layers before these dense layers, however, are usually very different from those of a standard ANN.                              

The most prominent types of layers are (a) \emph{convolutional layers}, (b) \emph{pooling layers}, and (c) \emph{dropout layers}: A convolutional layer usually consists of a small set of filters (e.g., 3x3, 5x5, \ldots), called \emph{kernels}, and convolves every input image with each of those kernels. Typically several such kernels are used as filters in a given convolutional layer to match desired shapes in the input images. This gives rise, for each kernel, to a new representation of the input image. These representations are called \emph{feature maps}. The convolutional layers are used with rectifiers to introduce non-linearity. A pooling layer decreases the number of parameters of the network by aggregating spatially-adjacent pixel values. One prominent type of pooling layers is max-pooling that replaces patches of an input feature map by the maximum value within each patch. While this reduces the sizes of the feature maps, it also makes the network more robust to small changes in the input data. Finally, a dropout layer randomly omits hidden units by setting their values to zero. Hence, the network cannot fully rely on them. This helps prevent overfitting.
 One usually resorts to one or more final dense layers based on a large number of nodes connected to the previous layer. Such layers are unlike the convolutional, pooling, and dropout layers, but the same as for traditional neural networks. It is the depth provided by these multiple layers, and the extensive mapping afforded by them that has given rise to the name {\it deep learning}. 
 
\begin{figure}
\centering
\subfigure[EW]{\includegraphics[width=2.5cm]{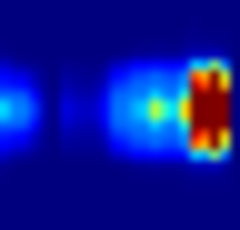}}
\hspace{1em}%
\subfigure[EA]{\includegraphics[width=2.5cm]{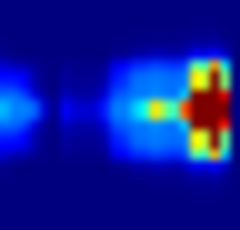}}
\hfill
\\
\hfill
\subfigure[RRab]{\includegraphics[width=2.5cm]{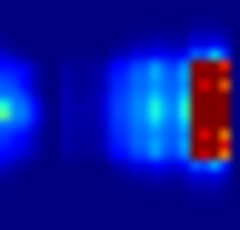}}
\hfill
\subfigure[RRc]{\includegraphics[width=2.5cm]{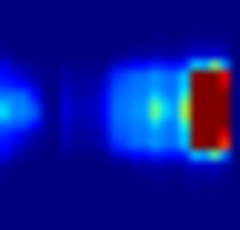}}
\hfill
\subfigure[RRd]{\includegraphics[width=2.5cm]{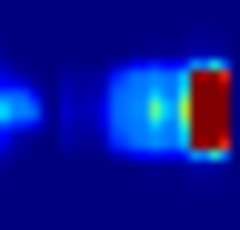}}
\hfill
\\
\subfigure[RS CVn]{\includegraphics[width=2.5cm]{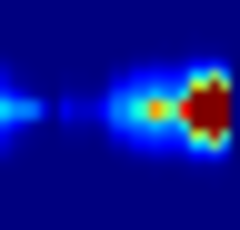}}
\hspace{1em}%
\subfigure[LPV]{\includegraphics[width=2.5cm]{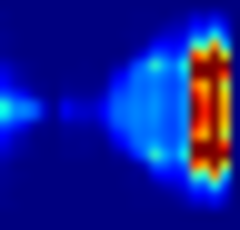}}
    \caption{Composite $dmdt$ images for all classes obtained by stacking all individual $dmdt$ images of each class. Given that we can visually discriminate between them, it should be clear that with purer base training samples image based classifiers will be able to classify them easily.}
    \label{fig:dmdt}
\end{figure}


\begin{table}
	\centering
	\caption{Accuracies for the multiclass classification (first row) and recall for the binary classifications (all other rows) are shown for both shallow and deep CNN on the fiducial $dmdt$-images. F1-score and Matthew's Coefficient are also provided for the deep CNN (binary classifications). Comparison with random forests with features is in the last column. Though not normally done for deep learning, we did 5 random train-test splits and report the range for the All7 networks and RF. Such a procedure may make more sense for smaller datasets.}
	\label{tab:fiducialsmallandlarge}
	\begin{tabular}{lccccc} 
\hline
Classes & $CNN_{shallow}$ & \multicolumn{3}{c}{$CNN_{deep}$} & RF\\
\cline{3-5}
&Recall & Recall &F1-score&MCC& Recall\\
\hline
All7 & $83.3\pm0.5$ & $83.2\pm0.3$ & - & - & $82.8\pm0.5$ \\
1/2 & 97/76 & 98/77.2 &0.97/0.81&0.79& 97/82 \\
1/4 & 99/67 & 98.7/66.3 &0.98/0.72&0.71& 99/63 \\
1/5 & 97/57 & 96.5/54.4 &0.94/0.63&0.58& 97/54\\
1/6 & 99/35 & 99.7/31 &0.99/0.41&0.44& 99/31 \\
1/8 & 99/23 & 99.6/20 &0.98/0.31&0.37& 99/0 \\
1/13 & 100/86 & 99.9/88.9 &0.99/0.93&0.93& 99/79\\
2/4 & 97/96 & 97.9/97.8 &0.98/0.97&0.95& 98/97 \\
2/5 & 96/99 & 97.3/99.4 &0.98/0.98&0.97& 98/99 \\
2/6 & 99/98 & 99.6/97.7 &1/0.97&0.97& 99/96 \\
2/8 & 97/92 & 98.2/89.7 &0.98/0.92&0.89& 97/92 \\
2/13 & 100/96 & 100/99 &1/1 &0.99& 99/96 \\
4/5 & 7/88 &58.6/94.3 &0.69/0.88&0.59& 66/95 \\
4/6 & 93/71 & 93.1/73.3 &0.94/0.7&0.64& 94/55 \\
4/8 & 92/91 & 95.1/84.9 &0.93/0.88&0.81& 93/88 \\
4/13 & 98/94 & 98.9/93.5 &0.99/0.94&0.93& 98/84 \\
5/6 & 99/9 & 97.6/12.37 &0.95/0.18&0.15& 99/21 \\
5/8 & 96/77 & 96.1/74.5 &0.94/0.76&0.70& 94/77 \\
5/13 & 100/88 & 99.8/96.8 &1/0.97&0.97& 99/89 \\
6/8 & 83/93 & 77.5/90.7 &0.74/0.92&0.66& 76/96 \\
6/13 & 100/97 & 98.9/98.1 &0.98/0.99&0.97& 96/91 \\
8/13 & 98/97  &99.3/93 &0.99/0.95&0.94& 98/91 \\
\hline
	\end{tabular}
\end{table}

For approaching the classification task at hand, we considered a multi-layer CNN instance, called \emph{deep}  network. Selecting good layers and parameters is, as yet, more an art than science. An interesting research direction is the use of Bayesian optimization for choosing the best network hyperparameters. As a first step towards this, we considered a far simpler \emph{shallow} network and were pleasantly surprised by its equally good performance for the base all-class classifications compared to the performance of the \emph{deep} network. The code detailing the structure of both networks is provided in Listing~\ref{code:cnn} and Listing~\ref{code:cnnsmall}, respectively. The listings include type and size of layers, number and size of kernels, and dropout fractions. We have used the theano framework (http://deeplearning.net/software/theano/) with lasagne (https://lasagne.readthedocs.io/en/latest/) for our runs. 

\section{Experimental Evaluation}
\label{sec:tests}

Light curves  were converted to $dmdt$-images, normalized as described in Sec.~\ref{sec:data}, and used as inputs to the network. We used 500 training epochs for the \textit{deep} network and 300 for the \textit{shallow} network with $20\%$ samples reserved for testing for both configurations. We used a learning rate of 0.0002 and the Adaptive Momentum (ADAM) algorithm to train all our models. All our models have been trained on a single NVIDIA GeForce GTX 560 graphics processing unit (GPU). It takes $\sim5.5$ and 42.3 seconds per epoch for the shallow and deep networks respectively when trained on the CRTS-N training set. Training RFs is one to two orders of magnitude faster - after computing the features. Depending on how complex the features are, the computing time can vary a lot. We used Red Hat linux release 6.6, python 2.7.2.

\subsection{Binary classification}
\label{subsec:bc}
We trained the network with pairs of classes as well as with all seven classes together. When used in binary mode we noticed poor performance when class 1 is involved (see Col.~2 of Table~\ref{tab:fiducialsmallandlarge}). It is not unexpected since class 1 contains two-thirds of all objects, and when paired with an individual class, it overwhelms every other class easily. 
Class 13 reached an accuracy of 89\% - the highest - against class 1. These are the Long period variables (LPVs) and the long-term structure is likely getting picked up. In general the separation of all other classes with class 13 was similarly far better than other binary comparisons. Except in a few cases, the binary $dmdt$-classifier did comparable or better than the corresponding feature-based random forests ($RF$) classifier (see Table~\ref{tab:fiducialsmallandlarge} and Sec.~\ref{sec:rf}).

\begin{minipage}{7cm}
\begin{lstlisting}[breaklines=true,language=python,frame=single,caption={Convolutional Neural Network - deep},    label={code:cnn},commentstyle={\color{gray}\fontseries{lc}\selectfont\itshape},columns=fullflexible]
layers = [
    InputLayer,
    Conv2DLayer(64, size:3x3, rectify),
    MaxPool2DLayer(2x2)),
    DropoutLayer(0.1),
    Conv2DLayer(128, size:5x5, rectify), 
    Conv2DLayer(256, size:5x5, rectify),
    DenseLayer(512),
    DropoutLayer(0.5),
    DenseLayer(512),
    DenseLayer(all, softmax),
]
\end{lstlisting}
\end{minipage}

\begin{minipage}{7cm}
\begin{lstlisting}[breaklines=true,language=python,frame=single,caption={Convolutional Neural Network - shallow network},    label={code:cnnsmall},commentstyle={\color{gray}\fontseries{lc}\selectfont\itshape},columns=fullflexible]
layers = [
    InputLayer,
    Conv2DLayer(32, size:3x3, rectify),
    DropoutLayer(0.1),
    DenseLayer(128),
    DropoutLayer(0.25),
    DenseLayer(128),
    DenseLayer(all, softmax),
]
\end{lstlisting}
\end{minipage}


\begin{table}
	\centering
	\caption{Experiments with varying dmdt}
	\label{tab:dmdtbins}
	\begin{tabular}{lc} 
\hline
Original binning (as outlined before):\\
$dm = \pm[0, 0.1, 0.2, 0.3, 0.5, 1, 1.5, 2, 2.5, 3, 5, 8]$ mags\\
$dt = [1/145, 2/145, 3/145, 4/145, 1/25, 2/25, 3/25, 1.5, 2.5, 3.5, 4.5,$\\
\hspace{15pt} $5.5, 7, 10, 20, 30, 60, 90, 120, 240, 600, 960, 2000, 4000]$ days\\
Average accuracy: 83\%\\
\hline
New 1:\\
$dm = \pm[0,0.05, 0.1, 0.15, 0.2, 0.25, 0.3, 0.4, 0.5, 1, 1.5, 2, 2.5, 3, 5, 8]$ mags\\
$dt = [1/145, 2/145, 3/145, 4/145, 1/25, 2/25, 3/25, 1.5, 2.5, 3.5, 4.5,$\\
\hspace{15pt} $5.5, 7, 10, 20, 30, 60, 90, 120, 240, 600, 960, 2000, 4000]$ days\\
Average accuracy: 84.5\%\\
\hline
New 2:\\
$dm = \pm[0,0.05, 0.1, 0.15, 0.2, 0.25, 0.3, 0.4, 0.5, 1, 1.5, 2, 2.5, 3, 5, 8]$ mags\\
$dt = [1/145, 2/145, 3/145, 4/145, 1.5, 2.5, 3.5, 4.5, 5.5, 7, 20,$\\
\hspace{15pt} $60, 120, 600, 960, 4000]$ days\\
Average accuracy: 84.3\%\\
\hline
New 3:\\
$dm = \pm[0, 0.1, 0.2, 0.3, 0.5, 1, 1.5, 2, 2.5, 3, 5, 8]$ mags\\
$dt = [1/145, 2/145, 3/145, 4/145, 1.5, 2.5, 3.5, 4.5, 5.5, 7, 20,$\\
\hspace{15pt} $60, 120, 600, 960, 4000]$ days\\
Average accuracy: 82.7\%\\
\hline
\end{tabular}
\end{table}

\begin{table}
	\centering
	\caption{Accuracy (row 1) and recall (other rows) for models trained using the shallow network on background subtracted images. (a) $CNN_c$: per class background, (b) $CNN$: pseudo-cadence background, (c) $CNN_{pmax}$ : max cadence background estimated from the training dataset subtracted from each $dmdt$ image in the dataset,(d) $CNN_e$: small $dm$ values eliminated (here 6 rows from our $dmdt$ model).}
	\label{tab:accuracies}
	\begin{tabular}{lcccc} 
\hline
Classes & $CNN_c$ & $CNN_p$ & $CNN_{pmax}$ & $CNN_e$ \\
\hline
All7 & 99.1 &83.2 &	 83.1  & 75.3 \\
1/2 & 99/100 & 98/75 &	 98/76 &98/48 \\
1/4 & 99/100 & 98/69&	 98/70 & 98/55\\
1/5 & 99/98 & 98/55&	 97/57 & 98/24\\
1/6 & 99/99 & 99/37&	 99/33 & 100/0\\
1/8 & 100/97 & 99/20&	 99/24 & 100/0\\
1/13 & 100/94 & 100/88&	 100/88 & 99/82\\
2/4 & 100/100 & 98/96&	 98/96 & 96/85\\
2/5 & 99/100 & 95/99&	 95/99 & 86/87\\
2/6 & 100/98 & 99/98&	 99/97 & 97/64\\
2/8 & 99/99 & 98/91&	 98/90 & 88/73\\
2/13 & 100/100 & 100/97& 100/96 & 99/91\\
4/5 & 100/100 & 71/91&	 68/91 & 65/91\\
4/6 & 100/100 & 91/77&	 96/52 & 94/56\\
4/8 & 99/100 & 92/91&	 94/90 & 91/83\\
4/13 & 100/98 & 99/93&98/93 & 97/87\\
5/6 & 100/98 & 98/16&  99/10 & 98/16\\
5/8 & 99/100 & 95/78&	 95/80 & 91/62\\
5/13 & 99/100 & 100/94&	 99/95 & 99/91\\
6/8 & 100/100 & 79/93&	 82/92 & 62/95\\
6/13 & 100/100 & 100/97&	100/97 &97/97 \\
8/13 & 99/100 & 98/97&	 98/100 & 98/93\\
\hline
	\end{tabular}
\end{table}


\subsection{Multi-class classification}
\label{subsec:mcc}

When used in the 7-class mode, the $dmdt$-images produced an average accuracy of $83\%$. This is remarkable in itself given the sparse nature of the data and no fine tuning of the parameters. The performance is comparable with feature-based $RF$-classifier (see Table~\ref{tab:fiducialsmallandlarge} last row, and Figs.~\ref{fig:confcnn} and \ref{fig:confrf}). Class 1 still dominates to an extent but not as blatantly as in the binary cases. It still leaves a lot to be desired if one wishes to use it in real-time for light curves containing far fewer points, and binary classification may be somewhat preferable. 

\subsection{Varying \textit{dmdt} binning}
\label{subsec:dmdtbins}

Input image size of 23x24 is small for CNNs and the training is relatively quick. As a result one can consider finer binning in both $dm$ and in $dt$. On the other hand, the discriminating structure likely resides in a few smaller areas, and one could use more granular binning. While it is desirable to determine the binning for a given survey and classes under consideration, such a systematic approach will need more extensive work. Here we report some quick experiments. We give below the variations we tried and the corresponding results (see Table~\ref{tab:dmdtbins}).

We notice that finer $dm$ bins improves performance a little, and fewer $dt$ bins do not seem to adversely affect the performance. Each experiment is time consuming and we continue our efforts to fine tune the parameters to get better results. As stated earlier the current results are already usable. Exploring these more is an obvious area of further advance.

\subsection{Background Subtraction}
\label{subsec:backgroundsub}

How can we improve the classification further? Individuals look at differences as well as similarities in order to classify objects. Consider spectra: given other common things, one asks if a particular line is too broad, or narrow, or missing, or extra-intense. We do the same when looking at light curves. With $dmdt$-images we have squared the number of points and distributed them over a rectangle. If we could remove an underlying, common background, the class-membership may become more obvious. It is a non-trivial task at best. We consider a few possibilities for determining such a background.

A $dmdt$-image can be said to be made of three components (1) a static background, $b$, that results primarily from the cadence of the survey. No matter what kind of astronomical object there is, one will always find pairs of observations with large $dt$ and small $dm$, and peaks at specific $dt$ depending on the survey cadence, (2) a more specific background related to the class-membership, $c_i$, for the $ith$ class. This is the kind of $dmdt$ one would get from a densely sampled prototype, and (3) something like an individual signature, $s$, for each object.

$$\hbox{{\it dmdt}--image} = b + c_i + s$$

We formulate the foreground-background separation problem by drawing parallels to video surveillance tasks. We interpret each $dmdt$-image as a video frame, vectorize it and then stack up all the $dmdt$-images columnwise to create a matrix $M$. We then decompose this matrix into a sum of a low rank matrix, $L$, and a sparse matrix, $S$. Each column of the low rank matrix corresponds to the background in the corresponding $dmdt$-image and each column of the sparse matrix corresponds to the foreground of the corresponding $dmdt$-image. $L$ and $S$ can be obtained by solving the following expression where L is forced to be low-rank and S is forced to be sparse:
\begin{center}
$ \underset{L,S}{\textbf{Min}} \|M - L - S\|_{2}$\\
\end{center}
We use the Robust PCA algorithm \cite{Netrapalli2014} for decomposing the matrix in this way. We use this particular method as it is one of the earliest method which is a provable non-convex algorithm in contrast to other works that rely upon convex relaxations of the actual objective.

\begin{figure}
\centering
\subfigure[Class 1 (EW)]{\includegraphics[width=2.7cm]{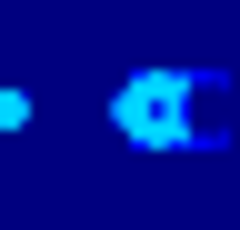}}
\hspace{1em}%
\subfigure[Class 2 (EA)]{\includegraphics[width=2.7cm]{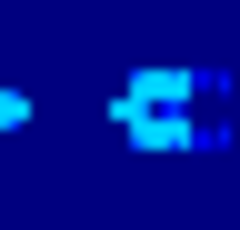}}
\\
\hfill
\subfigure[Class 4 (RRab)]{\includegraphics[width=2.7cm]{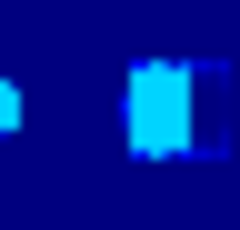}}
\hfill
\subfigure[Class 5 (RRc)]{\includegraphics[width=2.7cm]{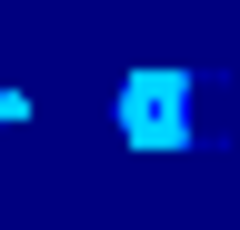}}
\hfill
\subfigure[Class 6 (RRd)]{\includegraphics[width=2.7cm]{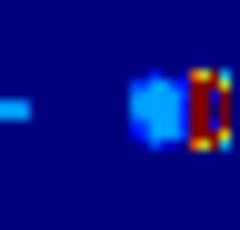}}
\hfill
\\
\subfigure[Class 8 (RS CVn)]{\includegraphics[width=2.7cm]{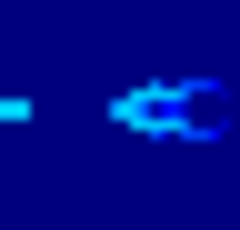}}
\hspace{1em}%
\subfigure[Class 13 (LPV)]{\includegraphics[width=2.7cm]{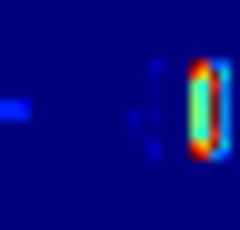}}
\caption{Class backgrounds determined using the robust PCA method. The $max$ for each class is shown.}
\label{fig:classback}
\end{figure}

\begin{figure}
\centering
\subfigure[Class 1 (EW)]{\includegraphics[width=2.7cm]{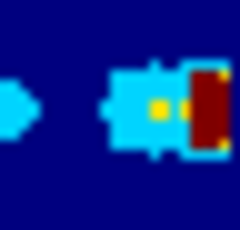}}
\hspace{1em}%
\subfigure[Class 2 (EA)]{\includegraphics[width=2.7cm]{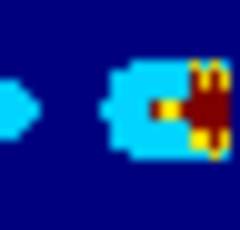}}
\\
\hfill
\subfigure[Class 4 (RRab)]{\includegraphics[width=2.7cm]{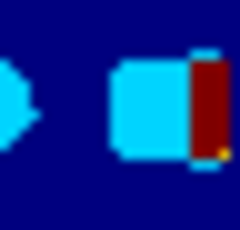}}
\hfill
\subfigure[Class 5 (RRc)]{\includegraphics[width=2.7cm]{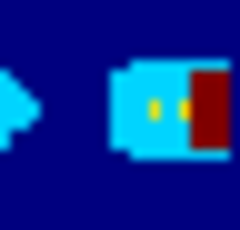}}
\hfill
\subfigure[Class 6 (RRd)]{\includegraphics[width=2.7cm]{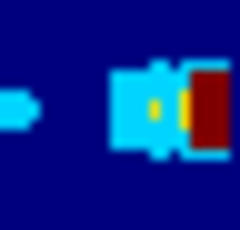}}
\hfill
\\
\subfigure[Class 8 (RS CVn)]{\includegraphics[width=2.7cm]{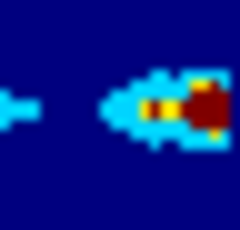}}
\hspace{1em}%
\subfigure[Class 13 (LPV)]{\includegraphics[width=2.7cm]{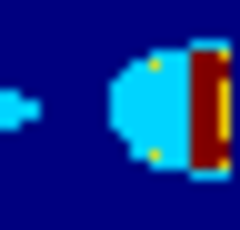}}
\caption{Median $dmdt$-images for individual classes.}
\label{fig:medians}
\end{figure}

We determine backgrounds to subtract before training in a few different ways: (1) for individual class backgrounds we consider $dmdt$-images of just that class to form the matrix $M$. Fig.~\ref{fig:classback} shows the class backgrounds. The class backgrounds are not like the median images of stacked class images (see Fig.~\ref{fig:medians}). The difference partly springs from non-uniform lengths of time-spans for individual light curves as well as differences in maximum brightness variations over different time-scales. We use the respective maximum background for each class. (2) For a pseudo-cadence background we consider all our objects together. We call this the pseudo-cadence background because all our objects in the current set are periodic variables. In order to not overwhelm this pseudo-background with a single class, we take 500 samples of each type (or all, if the training sample has fewer than 500). (3) As another possibility we ignored the class imbalance, took all training samples, and used the $max$ from the background for subtraction from all training and testing samples. (4) For a true background we will need to consider $dmdt$-images for light curves of randomly selected objects (or perhaps a large number of standard stars - stars known not to vary). Since a vast majority of these objects will be constant within error-bars over the entire time-span, the corresponding $dmdt$-images will consist of a thin line along the $dm = 0$ midrib. After the backgrounds are subtracted, the CNN is trained on the foreground images. 

Subtracting the class background provides far better results as expected (see Table~\ref{tab:accuracies}) since we use class information to subtract a specific background even during testing, and in the real world we are not be privy to this information. However it does show that removing appropriate background accentuates the different features between different classes. However. the removal of the pseudo-cadence background is somewhat worse than not removing any background. We mimicked cadence background removal by blanking the middle 2,4,6 rows of the $dmdt$-images, but they did not provide better results than not removing the background either. Clearly, better modeling of the background is required, a more time-consuming job that we will be taking up in the near-future.


\subsection{Transfer Learning}
\label{subsec:transferlearning}

One of the real power of the $dmdt$-technique is its cross-survey applicability. We used models trained with the CRTS-N $dmdt$-images and tested them on CRTS-S $dmdt$-images with same classes \cite{Drake2017}, but no overlapping objects, and with PTF $dmdt$-images with a subset of the same objects as in the CRTS-N sample. CRTS-S uses the same asteroid-finding cadence as CRTS-N and also has an open filter. PTF used a more mixed cadence with a greater emphasis on looking for explosive events including a repeat cadence of 1, 3, 5 nights. We used PTF data taken with the $r$ filter. The results using both the shallow and deep CNNs are given in Table~\ref{tab:transferlearning}. The numbers are not as good as with CRTS-N, but that is not unexpected. In fact, for many classes, especially for CRTS-S, they are better than one would naively expect. In case of PTF the survey cadence is very different in addition to the aperture and wavelength range and the results are somewhat worse. But the very fact that they are still usable, and definitely a good starting point indicates the merit of using such a technique. With proper survey-based background subtraction the results should improve further. The implications for domain adaptation are obvious, especially with applicability to forthcoming surveys like ZTF and LSST.


\begin{table}
	\centering
	\caption{Accuracies (row 1) and recall (other rows) for models trained on $dmdt$-images from CRTS-N and tested on CRTS-S, and PTF for the CNN shallow and deep networks.}
	\label{tab:transferlearning}
	\begin{tabular}{lcccc} 
\hline
Class & \multicolumn{2}{c}{$CRTS_S$} & \multicolumn{2}{c}{PTF} \\
\hline
 & shallow & deep & shallow & deep \\
\hline
All7 & 69.5 & 69.9 & 66.5 & 66\\
1/2 & 98/70&98/69 &95/44 &95/45  \\
1/4 & 99/49&98/52 & 95/38&96/32\\
1/5 & 99/23&97/37 & 98/8&97/13\\
1/6 & 99/13&99/13 &99/4 &98/23\\
1/8 & 99/3&99/4& 86/63&88/48\\
1/13 & 99/82&99/82 & 75/83&64/80\\
2/4 & 96/97&95/98 & 85/94&80/96\\
2/5 & 96/98& 95/99& 70/94&73/96\\
2/6 & 99/92& 99/96& 99/51&99/75\\
2/8 & 97/81& 97/76& 28/97&55/94\\
2/13 & 99/95 & 99/96& 85/76&61/83\\
4/5 & 69/90& 66/90& 80/59&57/85\\
4/6 & 88/72& 94/53& 98/11&97/8\\
4/8 & 82/86& 92/62& 41/97&72/91\\
4/13 &92/93 &95/93 & 70/89&79/97\\
5/6 & 99/1&97/6 & 99/0&93/5\\
5/8 & 96/51& 96/44& 46/92&80/53\\
5/13 & 99/84& 99/91& 84/83&81/89\\
6/8 & 67/89& 75/88& 3/99&15/97\\
6/13 &96/94 &94/95 & 80/89&78/85\\
8/13 &98/91 & 98/90& 77/67&58/89\\
\hline
	\end{tabular}
\end{table}



\section{Discussion}
\label{sec:discussion}

We have shown how to transform light curves to simple $dmdt$-images for use with canned as well as simpler CNNs for out-of-the-box classification of objects with performance comparable to random forests, and without having to resort to designing or extracting features, or other necessary evils like dimensionality reduction. The internal features the CNN uses need to be explored further using tools like deconvolutional networks. That will make the results interpretable, and provide insights. We have also shown various paths to take in order to improve the results further e.g. background subtraction and varying the $dmdt$ bins. We have further demonstrated the application of the technique to transfer learning and thereby classifying objects from a completely different survey. 

\subsection{Comparison with RF}
\label{sec:rf}

Random forests (RF) tend to be very versatile and difficult to beat in performance. Hence we compare the performance of our technique with random forests. We use the features given in Table~\ref{tab:rffeatures} for our RF setup. The output is shown in  Table~\ref{tab:fiducialsmallandlarge}. The features we have used are generic, and designer features would provide better recall and precision for select classes. 

We find the performance of the shallow CNN comparable to that of unweighted RF. In a way this is remarkable since it is akin to providing the light curve almost in its raw format and getting a  classification. For some classes the recall is low, possibly due to the sparseness of the light curves. RFs have provided better results when used with features based on non-sparse light curves \cite{Richards2011,Dubath2011}. Combining the features with the CNN to form a deep-and-wide network will likely provide better performance than either.

\begin{table}
\centering
\caption{Random forest features. The first three are not used in RF whereas the remaining 18 are fairly generic features often used in classification e.g. \cite{Richards2011,Donalek2013,Graham2014}. Formulae for the features are from $http://nirgun.caltech.edu:8000/scripts/description.html\#method\_summary$}.
\label{tab:rffeatures}	
\begin{tabular}{lc} 
		\hline
		Feature & Formula \\
		\hline
        meanmag & $<mag>$ \\
minmag & $mag_{min}$ \\
maxmag & $mag_{max}$ \\
amplitude & $0.5 * (mag_{max} - mag_{min})$ \\
beyond1std & $p(|(mag - <mag>)| > \sigma)$ \\
flux percentile ratio mid20 & $(flux_{60} - flux_{40}) / (flux_{95} - flux_{5})$ \\
flux percentile ratio mid35 & $(flux_{67.5} - flux_{32.5}) / (flux_{95} - flux_{5})$ \\
flux percentile ratio mid50 & $(flux_{75} - flux_{25}) / (flux_{95} - flux_{5})$ \\
flux percentile ratio mid65 & $(flux_{82.5} - flux_{17.5}) / (flux_{95} - flux_{5})$ \\
flux percentile ratio mid80 & $(flux_{90} - flux_{10}) / (flux_{95} - flux_{5})$ \\
linear trend & b where mag = a * t + b \\
max slope & $max(|(mag_{i+1}-mag_{i})/(t_{i+1}-t_{i})|)$ \\
median absolute deviation & $med(flux - flux_{med})$ \\
median buffer range percentage & $p(|flux - flux_{med}| < 0.1 * flux_{med})$ \\
pair slope trend & $p(flux_{i+1} - flux_{i} > 0; i = n-30, n)$ \\
percent difference flux percentile & $(flux_{95} - flux_{5}) / flux_{med}$ \\
skew & $\mu_3/\sigma^3$ \\
small kurtosis & $\mu_4/\sigma^4$ \\
std & $\sigma$ \\
stetson j & $var_j$ (mag) \\
stetson k & $var_k$ (mag) \\
		\hline
	\end{tabular}
\end{table}
\begin{figure}
\centering
	\includegraphics[width=6cm]{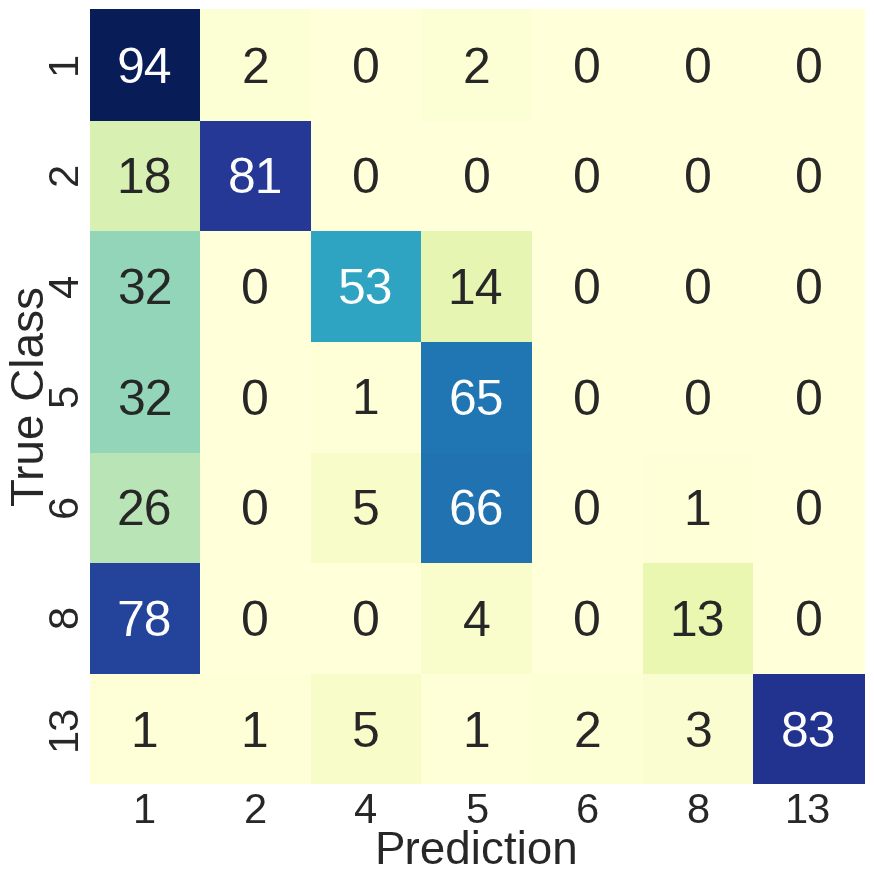}
    \caption{Confusion matrix for shallow CNN with fiducial $dmdt$-images. Note the high misclassification between classes 5 and 6, both RR Lyrae.}
    \label{fig:confcnn}
\end{figure}
\begin{figure}
\centering
	\includegraphics[width=6cm]{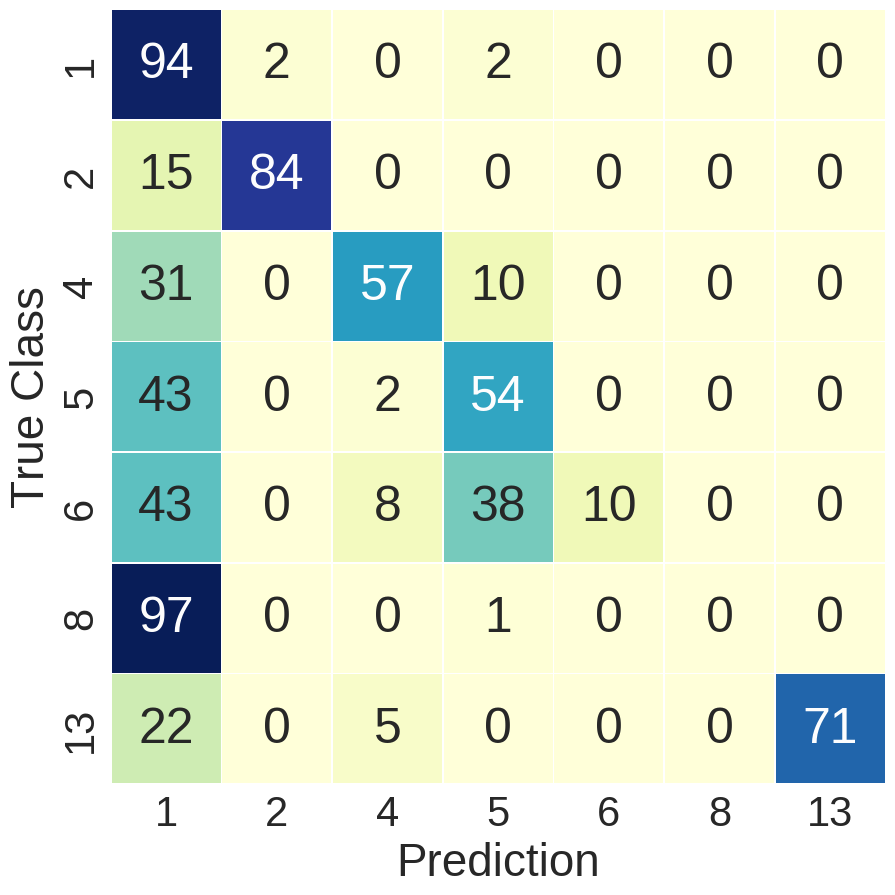}
    \caption{Confusion matrix for random forest using features. The numbers are given as percentages. Overall classification accuracy is 82.2\%. Yellow, Green, and Blue indicate successively larger percentages from 0 to 100.}
    \label{fig:confrf}
\end{figure}

\subsection{Misclassified Sources}
\label{sec:misclassified}

\begin{figure}
\centering
\subfigure{\includegraphics[width=6cm]{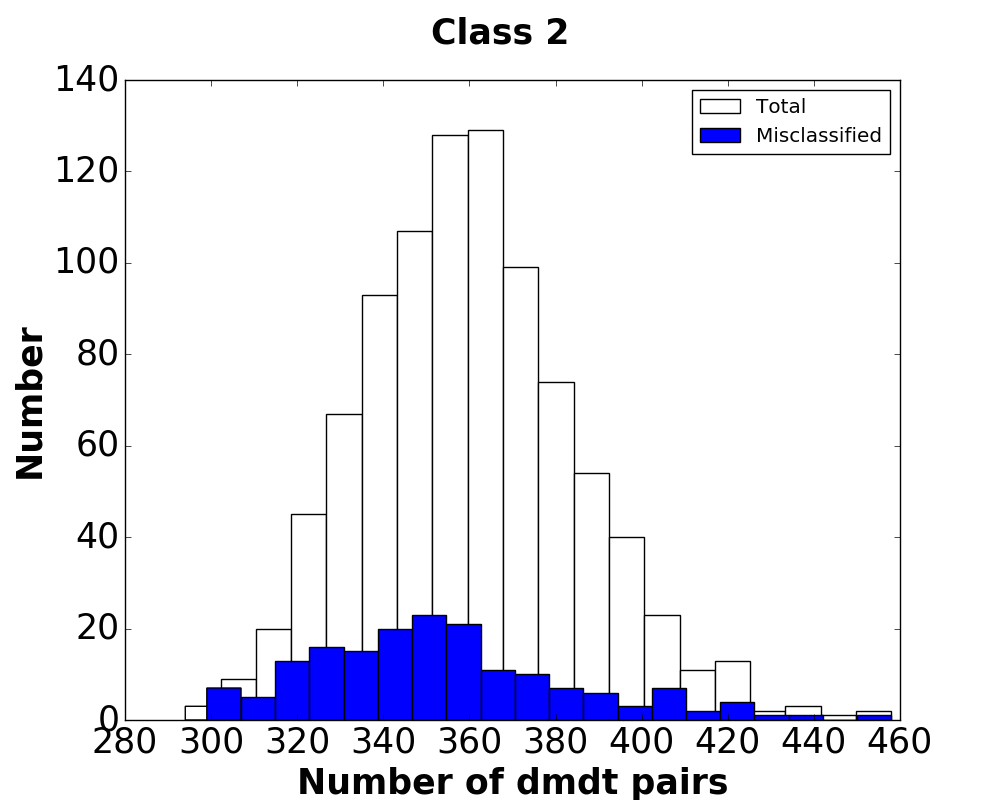}}
\hfill
\subfigure{\includegraphics[width=6cm]{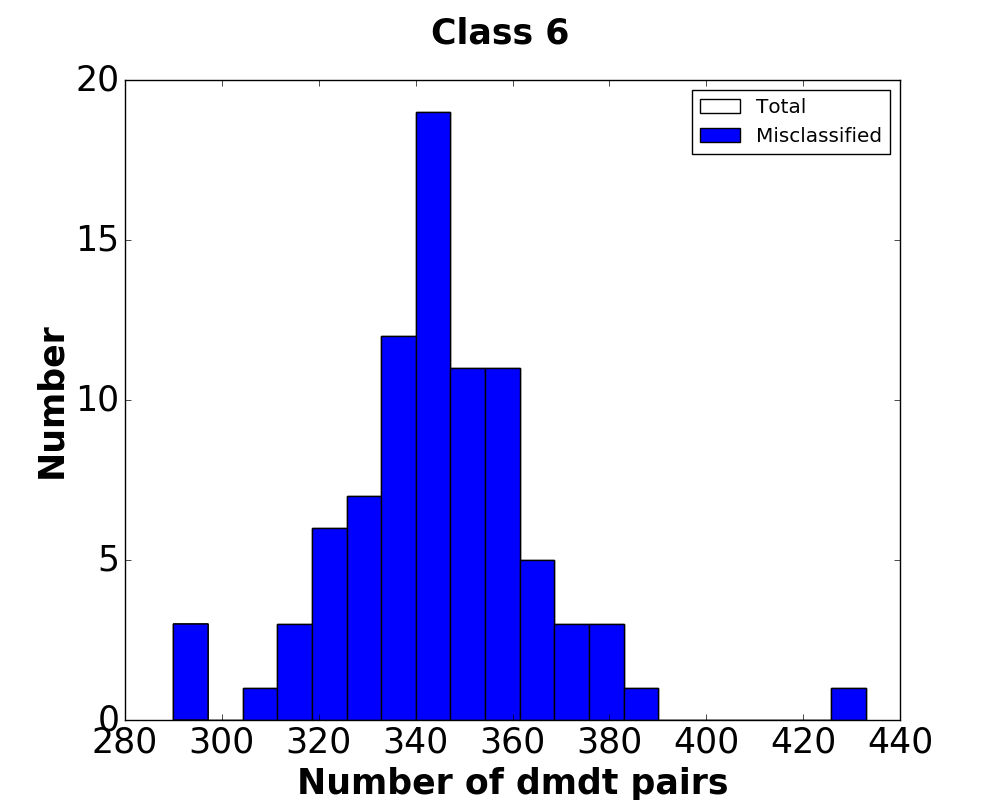}}
\caption{Histograms of misclassified objects for classes 2 (top) and 6 (bottom). All class 6 (RRd) objects were misclassified, two-thirds of them being classified as another type of RR Lyrae (RRc). 19\% of class 2 (EA) objects were misclassified, most of them as class 1 (EAs), and the distribution of points in the misclassified set suggests that though there is a slight trend, it is not the shortest light curves that were thus misclassified. That is true in general for misclassifications in other classes as well. }
\label{fig:misclass}
\end{figure}

\begin{figure}
\centering
\subfigure[Foreground 1]{\includegraphics[width=2.7cm]{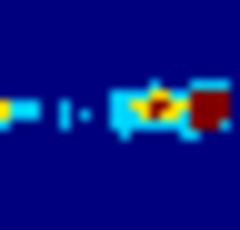}}
\hspace{1em}%
\subfigure[Foreground 2]{\includegraphics[width=2.7cm]{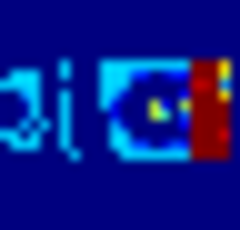}}
\caption{Two types of foregrounds were seen for class 1, suggesting a possible split in the dataset. This dichotomy needs to be investigated further.}
\label{fig:class1foreg}
\end{figure}

The confusion matrices (Figs.~\ref{fig:confcnn} and \ref{fig:confrf}) during our various experiments showed that for some classes large fractions of objects were misclassified (see Fig.~\ref{fig:misclass}). We investigated the light curves for some of these sources in order to identify the source of errors. In some cases it was a genuine error (wrong label) indicating that the network was working well. In some other cases the misclassification was owing to a sparse light curve indicating that in a handful of cases a smaller number of features may be tilting the classifications one way or another. In still other cases, the subclasses were just too close for the technique to discern them apart just from the $dmdt$-images based upon the light curves (e.g. RR Lyrae of different types). In the case of EW and EA classes we suspect that the technique may be teasing apart subclasses (e.g. based on separation) or geometric dependence. For example, we noticed two distinct kinds of foregrounds (see Fig.~\ref{fig:class1foreg}), and this needs to be investigated further.


\subsection{Future Work}
\label{sec:futurework}

We will explore various possibilities related to varying CNN hyperparameters, improving background subtraction for more reliable classification, expanding to more classes and surveys, as well as identifying the misclassified sources. We will also experiment to make the technique more useful in the real-time cases with far fewer data points. We did a couple of tests using error-bars to augment smaller classes, but that did not work well. That needs to be explored further for reducing the unbalancedness of the different classes. Also there is the possibility of using Generative Networks to create large simulated examples for different classes for understanding the features that really separate different classes. The number of possibilities is large -- we invite others to explore them as well.

\section*{Acknowledgements}

This work, and CRTS survey, was supported in part by the NSF grants AST-0909182, AST-1313422, AST-1413600, and AST-1518308, and by the Ajax Foundation. KS thanks IIT-Gandhinagar and the Caltech SURF program.






\bibliographystyle{IEEEtran}
\bibliography{main}



\end{document}